\pdfoutput=1

\documentclass[twocolumn]{aastex62}

\usepackage{cleveref}

\usepackage{graphics, graphicx}

\newcommand\cone{$3MM$-1}
\newcommand\ctwo{$3MM$-2}

\begin{document}

\title{Discovery of a dark, massive, ALMA-only galaxy at $z\sim5-6$ in a tiny 3-millimeter survey }

\author[0000-0003-2919-7495]{Christina C. Williams}
\affiliation{Steward Observatory, University of Arizona, 933 North Cherry Avenue, Tucson, AZ 85721, USA}\affiliation{NSF Fellow}

\author[0000-0002-2057-5376]{Ivo Labbe}
\affiliation{Centre for Astrophysics \& Supercomputing, Swinburne University of Technology, PO Box 218, Hawthorn, VIC 3112, Australia}

\author[0000-0003-3256-5615]{Justin Spilker}
\affiliation{Department of Astronomy, University of Texas at Austin, 2515 Speedway, Stop C1400, Austin, TX 78712, USA}

\author[0000-0001-7768-5309]{Mauro Stefanon}
\affiliation{Leiden Observatory, Leiden University, NL-2300 RA Leiden,Netherlands}

\author[0000-0001-6755-1315]{Joel Leja}
\affiliation{Harvard-Smithsonian Center for Astrophysics, 60 Garden St. Cambridge, MA 02138, USA}\affiliation{NSF Fellow}

\author[0000-0001-7160-3632]{Katherine Whitaker}
\affiliation{Department of Physics, University of Connecticut, 2152 Hillside Road, Unit 3046, Storrs, CT 06269, USA}

\author[0000-0001-5063-8254]{Rachel Bezanson}
\affiliation{Department of Physics and Astronomy and PITT PACC, University of Pittsburgh, Pittsburgh, PA, 15260, USA}

\author[0000-0002-7064-4309]{Desika Narayanan}
\affiliation{Department of Astronomy, University of Florida, 211 Bryant Space Science Center, Gainesville, FL 32611, USA}

\author[0000-0001-5851-6649]{Pascal Oesch}
\affiliation{Department of Astronomy, University of Geneva, 51 Ch. des Maillettes, 1290 Versoix, Switzerland  }\affiliation{International Associate, Cosmic Dawn Center (DAWN) at the Niels Bohr Institute, University of Copenhagen and DTU-Space, Technical University of Denmark   }

\author[0000-0001-6065-7483]{Benjamin Weiner}
\affiliation{Steward Observatory, University of Arizona, 933 North Cherry Avenue, Tucson, AZ 85721, USA}


\begin{abstract}

We report the serendipitous detection of two 3 mm continuum sources found in deep ALMA Band 3 observations to study intermediate redshift galaxies in the COSMOS field. One is near a foreground galaxy at 1\farcs3, but is a previously unknown  dust-obscured star-forming galaxy (DSFG) at probable $z_{CO}=3.329$, illustrating the risk of misidentifying shorter wavelength counterparts. The optical-to-mm spectral energy distribution (SED) favors a grey $\lambda^{-0.4}$ attenuation curve and results in significantly larger stellar mass and SFR compared to a Calzetti starburst law, suggesting caution when relating progenitors and descendants based on these quantities. The other source is missing from all previous optical/near-infrared/sub-mm/radio catalogs (``ALMA-only''), and remains undetected even in stacked ultradeep optical ($>29.6$ AB) and near-infrared ($>27.9$ AB) images. Using the ALMA position as a prior reveals faint $SNR\sim3$ measurements in stacked IRAC 3.6+4.5, ultradeep SCUBA2 850$\mu$m, and VLA 3GHz, indicating the source is real. The SED is robustly reproduced by a massive $M^*=10^{10.8}$M$_\odot$ and $M_{gas}=10^{11}$M$_\odot$, highly obscured $A_V\sim4$, star forming $SFR\sim300$ M$_{\odot}$yr$^{-1}$ galaxy at redshift $z=5.5\pm$1.1. The ultrasmall 8 arcmin$^{2}$ survey area implies a large yet uncertain contribution to the cosmic star formation rate density CSFRD(z=5) $\sim0.9\times10^{-2}$ M$_{\odot}$ yr$^{-1}$ Mpc$^{-3}$, comparable to all ultraviolet-selected galaxies combined. These results indicate the existence of a prominent population of DSFGs at $z>4$, below the typical detection limit of bright galaxies found in single-dish sub-mm surveys, but with larger space densities $\sim3 \times 10^{-5}$ Mpc$^{-3}$, higher duty cycles $50-100\%$, contributing more to the CSFRD, and potentially dominating the high-mass galaxy stellar mass function.

\end{abstract}

\section{Introduction} \label{sec:intro}
In past decades, single dish sub-millimeter surveys have identified populations of massive, dusty star-forming galaxies at $z>1$ \citep[e.g.][]{Casey2014}. While these galaxies are rare even at  
Cosmic Noon ($1<z<3$) when the star-formation activity in the Universe peaks, their contribution  
to the cosmic star-formation rate density (CSFRD) equals that of all optical and near-infrared selected galaxies combined 
\citep{MadauDickinson2014}. However, at $z>3$ the situation is much less certain. A tail of sub-millimeter selected galaxies have been confirmed beyond $z>4$, but they trace only the very tip of the star-formation rate (SFR) distribution at early times \citep[e.g.][]{Cooray2014,Strandet2017, Marrone2018}.  The total contribution of dust-obscured star-formation, and therefore the census of star-formation in the early Universe, is unknown.
Despite the strongly negative k-correction allowing sources to be found to $z=10$, the overwhelming majority of (sub)-mm selected galaxies continue to be confirmed at $z<3$ with  Atacama Large Millimeter/submillimeter Array (ALMA) spectroscopy \citep{Danielson2017, Brisbin2017}. 
While some dusty galaxies have been discovered beyond $z>5$ through gravitational lensing \citep{Spilker2016, Zavala2018Nat}, the lensing correction and selection effects make it challenging to establish their contribution to the CSFRD. 
Progress is hampered by the limited sensitivity and low spatial resolution of single-dish sub-mm observations and the difficulty of associating detections with counterparts in the optical-NIR.
Ultradeep SCUBA surveys over moderate $\sim100$ arcmin$^2$ are now pushing into the range of ``normal'' star formation rates (several 100 M$_\odot$/yr, main sequence galaxies) \citep[e.g.,][]{Koprowski2016,Cowie2017,Cowie2018} and extending to $z>4$, but the analysis is often limited by the ability to identify counterparts at other wavelengths and derive accurate redshifts.

ALMA has opened an avenue to address this issue through surveys at 
superior sensitivity and spatial resolution. ALMA deep fields at $\sim$1 mm  \citep{Dunlop2017, Aravena2016a} have probed to extremely deep flux limits over small areas ($<5$ arcmin$^2$). Progress has still been limited likely because dust obscured star formation preferentially occurs in massive galaxies \citep{Whitaker2017}, which are clustered and relatively rare ($\sim$0.1 arcmin$^{-2}$ at $\log(M/M_\odot)>$10.8 and $z\sim4$; \citealt{Davidzon2017}). Wider (10's arcmin$^2$) and shallower ($\sim$100-200$\mu$Jy) ALMA surveys at $\sim$1mm \citep{Hatsukade2018, Franco2018} are now approaching large enough areas to identify tentative massive candidates at $z>4$ \citep[e.g.][]{Yamaguchi2019}.

A promising development is to select at longer wavelengths ($>2$ mm), which optimizes the selection to dusty star formation at redshift $z>4$ \citep[][]{Bethermin2015b, Casey2018b}. However, the current state-of-the-art 
ALMA Spectroscopic Survey (ASPECS) at 3-mm  \citep{GonzalezLopez2019}, still only covered $\sim5$ arcmin$^{2}$, and identified 6 continuum sources, all at $z<3$.
Larger archival studies of ALMA 3 mm observations to find high-redshift candidates report some spectroscopic confirmations, but like the 1-mm redshift distribution, the majority lie at $z<3$ \citep{Zavala2018_3mm}. 
Recent advances with IRAM/GISMO provide evidence that $2$mm surveys favor selecting higher-redshift sources \citep{Magnelli2019}, although counterpart identification continues to be problematic due to large beam sizes.  Overall, the number of strong candidates for dust-obscured sources at $z>4$ remains small and as a result, the contribution of dust-obscured star formation in the early universe is poorly constrained.

\begin{figure*}
\includegraphics[scale=.23]{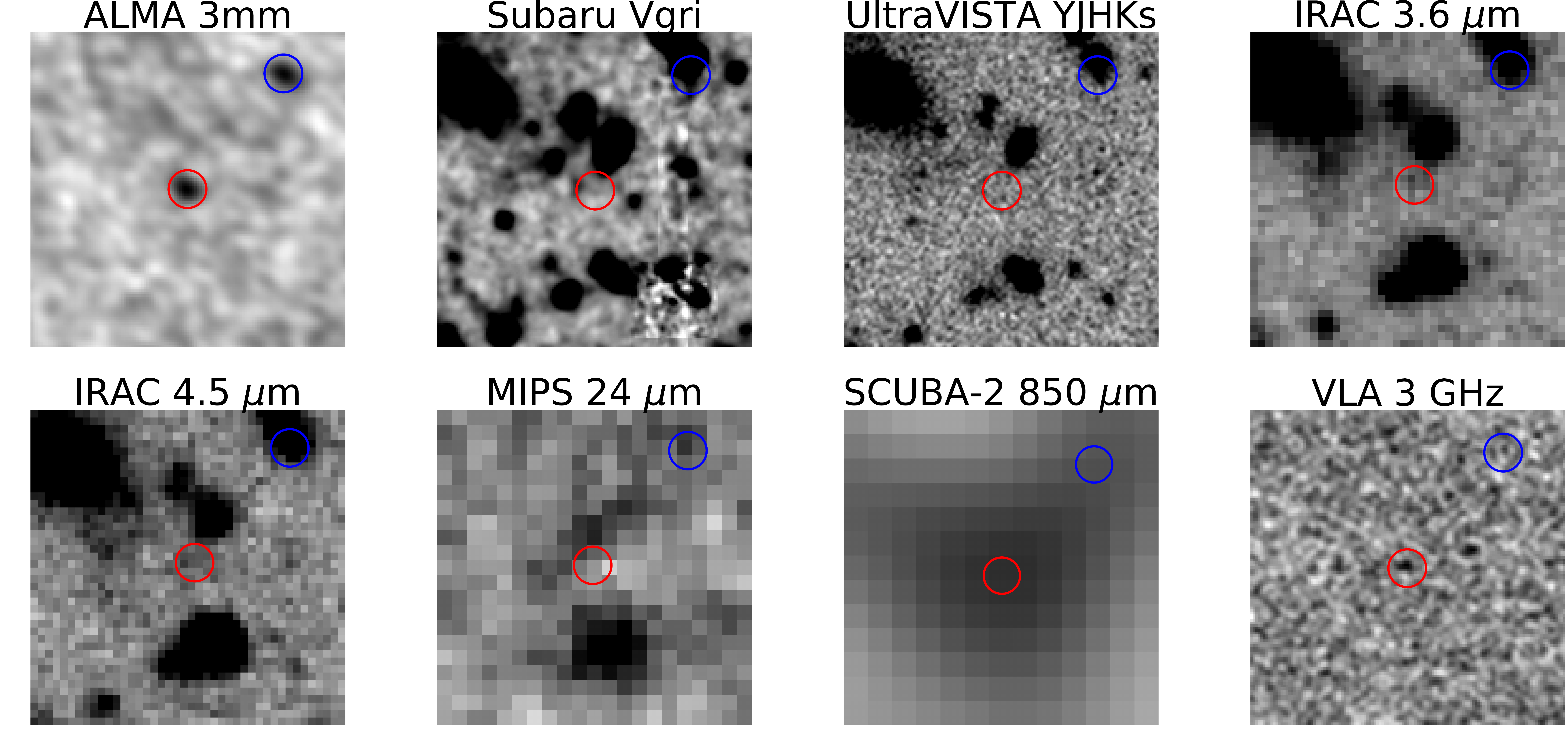}
\caption{ Cutouts (25$^{\prime\prime}$x25$^{\prime\prime}$) centered at 3-mm position of \cone\ (red circle; 3$^{\prime\prime}$ diameter).  \cone\ was not previously detected ($>$3$\sigma$) at any shorter wavelength, including deep optical and near-IR stacks, Spitzer, {\it Herschel}, and S2COSMOS SCUBA2 850$\mu$m. Remeasuring with the ALMA position as prior reveals marginal $2-3\sigma$ measurements in IRAC 3.6+4.5 and 850$\mu$m, consistent with heavy dust obscuration at $z>4$. It is faintly detected at 3GHz (4$\sigma$) indicative of a moderate radio excess due to a possible AGN. The 1.4 GHz image is excluded because neither source is significantly detected.  \ctwo\ is also identified (blue circle), and is blended with a foreground galaxy at $z=0.95$, 1.3'' North \citep{Muzzin2013,Laigle2016}. }
\label{cutout}
\end{figure*}

\begin{figure*}
    \centering
    \includegraphics[scale=0.6]{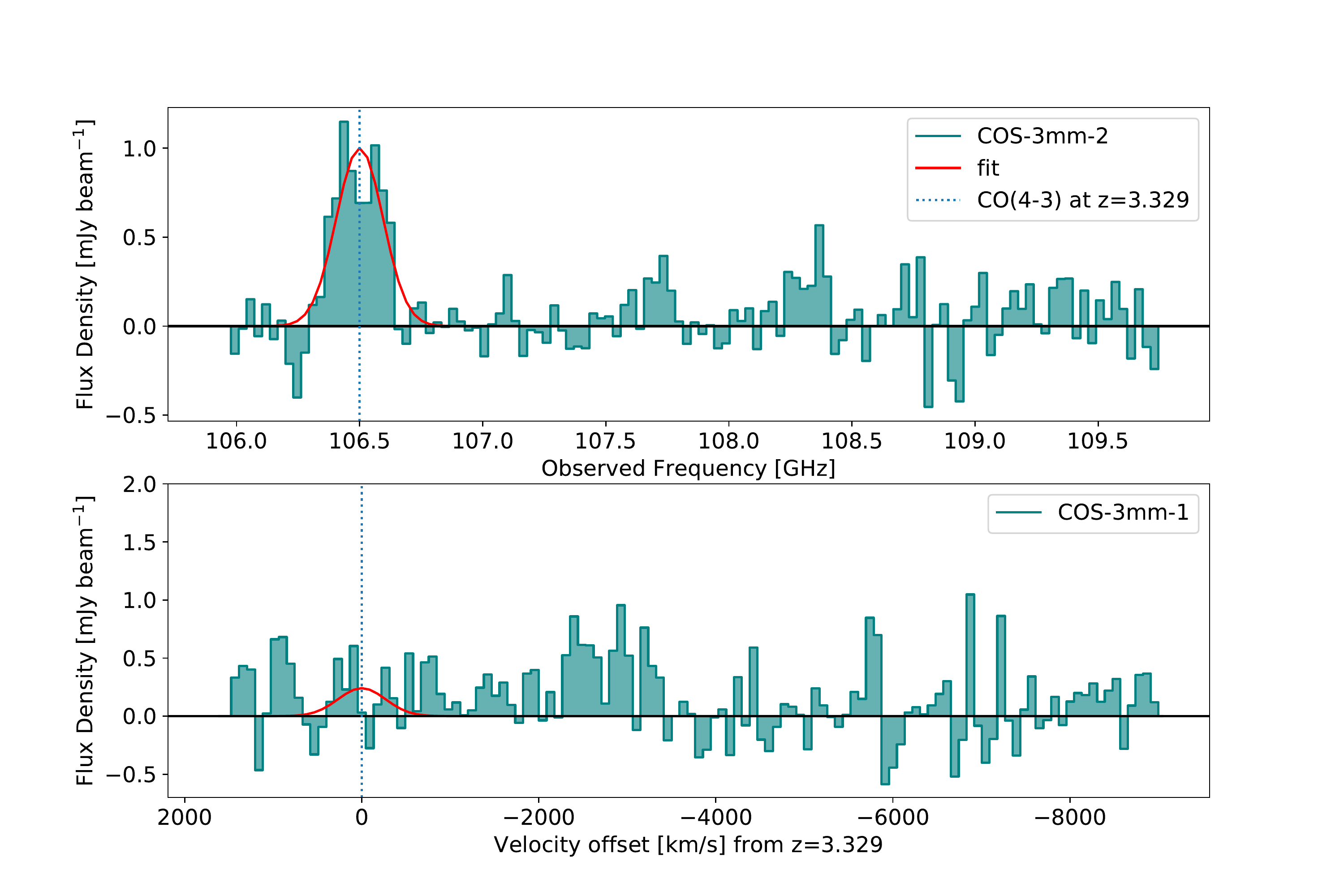}
    \caption{Portion of the observed ALMA Band 3 spectrum covering the detected CO line in \ctwo\ at 106.5 GHz (upper side band). The CO solution corresponding to CO(4-3) at z=3.329 is in excellent agreement with the photometric redshift measured in Section \ref{sec:sed}. The line flux is 0.66 $\pm$ 0.1 Jy km s$^{-1}$ and a Gaussian fit produces a width of $650$ km/s. No line is found at the same frequency in the spectrum of \cone\ (lower panel) with formal $SNR=0.8$. No lines are detected in the lower side band from $94 < \nu < 96.8$ GHz. }
    \label{fig:CO43}
\end{figure*}

\begin{figure*}
\includegraphics[scale=.5,trim= 20 150 20 0,clip]{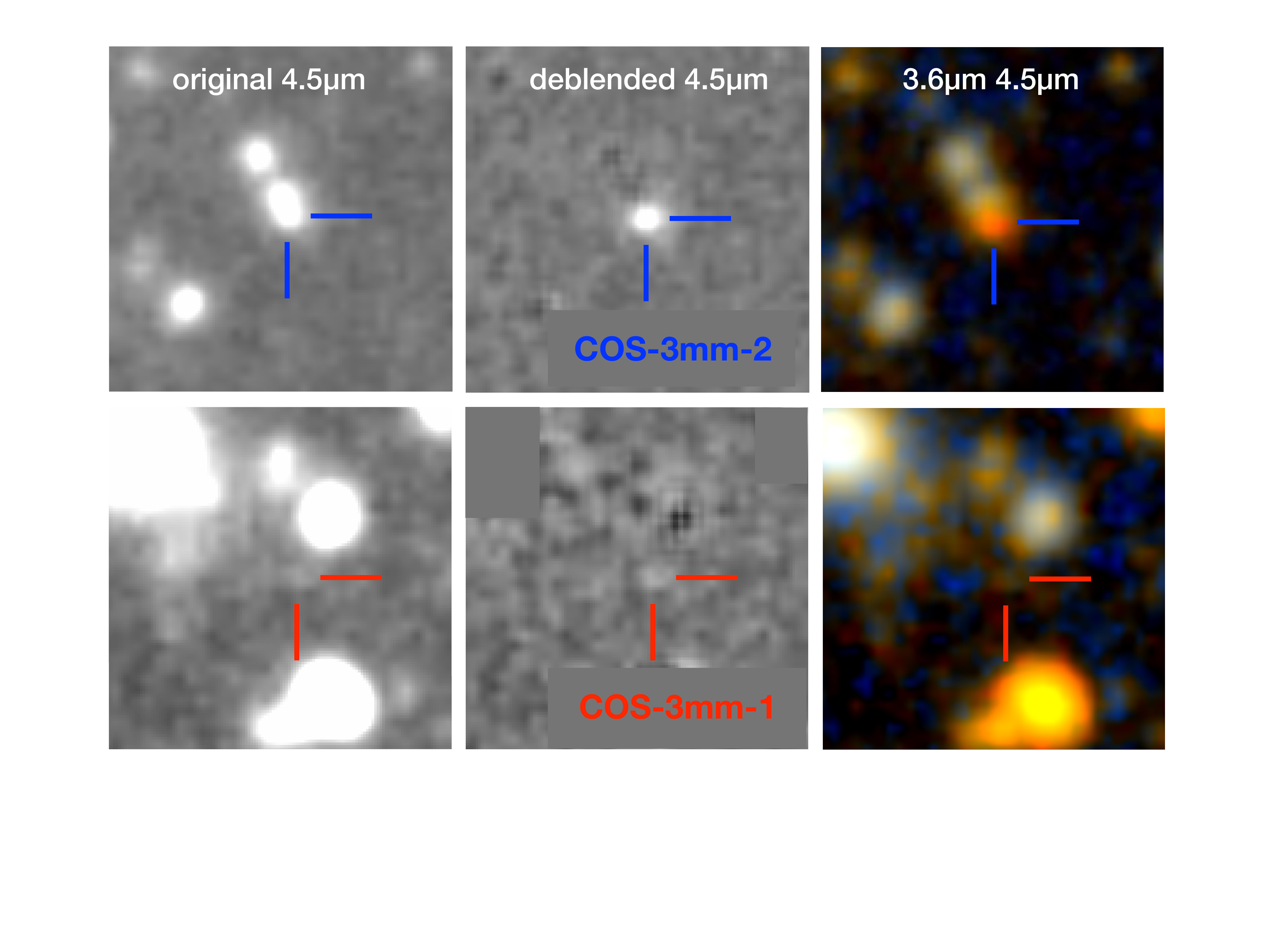}
\caption{Illustration of the deblending procedure using the \textsc{mophongo} software \citep{Labbe2015}. Deblending is performed by simultaneously fitting the pixels of all sources using the deep optical images and the ALMA positions as priors and accounting for differences in the PSFs. Top row: Deblending results for the 4.5$\mu$m band centered on the ALMA position of \ctwo\ ($12''\times12''$). Left panel shows the original 4.5$\mu$m image, middle panel shows \ctwo\ after other modeled sources have been subtracted, and the right panel a 3.6$\mu$m$-4.5\mu$m color image clearly indicating a vastly different IRAC color at the location of the ALMA source. Bottom row: Higher contrast, zoomed-in panels 
centered on the ALMA position of \cone, showing a faint $AB\sim25.2$ IRAC source after subtracting the PSF wings of a bright foreground neighbor.
}
\label{deblend}
\end{figure*}

Here we report the serendipitous discovery of 2 previously unknown sources in deep ALMA 3-mm observations in the COSMOS field, and assess the implications for dust obscured star formation at $z>3$. 
We assume a $\Lambda$CDM cosmology with  H$_0$=70 km s$^{-1}$ Mpc$^{-1}$, $\Omega_M$ = 0.3, $\Omega_\Lambda$ = 0.7, and a \citet{Kroupa2001} initial mass function (IMF).

\section{Methods} \label{sec:methods}

\subsection{ALMA millimeter interferometry}\label{sec:alma}
The ALMA observations are part of a program targeting CO(2-1) line emission in $z\sim1.5$ galaxies (2018.1.01739.S, PI: Williams; 2015.1.00853.S, PI: Bezanson) in the COSMOS field.  The data presented here include ALMA maps of five galaxies, and results on these targets are presented elsewhere (\citealt[][Williams et al. in prep]{Bezanson2019}). In one of these pointings we identified two serendipitous continuum sources, unrelated to the original target, which we will describe now in detail. 

ALMA Band 3 observations were carried out in two observing blocks on 2018 December 23 and 24 under program 2018.1.01739.S (PI: Williams). One 1.875 GHz spectral window was centered at the sky frequency (94.92 GHz) with 7.8 MHz ($\sim$24 km s$^{-1}$) channelization. Three additional 1.875\,GHz bandwidth spectral windows were placed at sky frequencies 96.8, 106.9, and 108.8 GHz for continuum observations, each with 15.6 MHz channelization. A total of 43 antennas were active reaching maximum baselines of 500m, for an angular resolution of $\sim$2 arcsec (synthesized beam major axis = 2.2'', minor axis = 1.7'' under natural weighting). The total time on-source was 97 min. 

The data were reduced using the standard ALMA Cycle 6 pipeline, and no problems with the pipeline calibration were found. The imaged data reach a continuum sensitivity of 5.7$\mu$Jy/beam at 101.9 GHz, and a typical line sensitivity of 55-65$\mu$Jy/beam per 100 km s$^{-1}$ channel. We imaged the data using natural weighting, creating a 101 GHz continuum image of the field from all four spectral windows. We imaged the data using 0.2'' pixels and created a 500$\times$500 pixel image, yielding images 100'' on a side. Given the ALMA primary beam at this frequency ($\sim$57'' FWHM), these images extend to approximately the 0.05 response point of the primary beam. 

\subsection{ALMA source detection} \label{sec:almadet}
Interferometric maps without correction for the primary beam response have uniform, normally-distributed noise properties across the field, and source detection significance is straightforward to measure from such maps. Two blind 3 mm continuum sources were apparent in this map, located 24.6'' and 38.2'' from the phase center, corresponding to primary beam response levels of 0.57 and 0.29, respectively. Each source is detected at a peak signal-to-noise ratio of $\sim8$; the probability of finding a Gaussian noise fluctuation of this magnitude given the number of independent beams in the images is exceedingly low ($<10^{-9}$). Both sources thus are real. The 3 mm flux densities of these sources, corrected for the primary beam response, are 155 $\pm$ 20 $\mu$Jy and 75 $\pm$ 10, hereafter referred to as \cone\ and \ctwo, respectively. 
Neither source is spatially resolved, based on a comparison of the peak pixel values and the integrated flux densities.
 
After finding both continuum sources in the combined map, we re-imaged the upper and lower sidebands of the data separately in order to determine the spectral index of each source at these frequencies. Thermal dust emission on the Rayleigh-Jeans tail has a very steep power-law index with $S_\nu \propto \nu^{2+\beta}$ and $\beta \sim 1.5-2$, while non-thermal synchrotron emission typically exhibits a negative spectral index, $S_\nu \propto \nu^{-0.8}$. Given their respective frequencies, we expect dust emission to be $\sim50-60$\% brighter in the upper sideband than the lower. We find spectral indices for both continuum sources in excellent agreement with the expectation for thermal dust emission.

We also re-imaged the data of both sidebands to search for blindly-detected emission lines in each of the two continuum sources, using channel widths ranging from 100-400 km s$^{-1}$.  \ctwo\ contains a serendipitous emission line centered at 106.5 GHz, with an integrated flux density of 0.66 $\pm$ 0.1 Jy km s$^{-1}$. The spectrum is shown in Figure \ref{fig:CO43}. A Gaussian fit indicates in a width of $630\pm 70$ km/s.  
Assuming the line is a transition of carbon monoxide, the possible redshifts are $z=$[0.08, 1.16, 2.25, 3.33, 4.41, 5.49]. We find no significant emission lines in the spectrum of \cone, and no evidence for a line at the same frequency of \ctwo. A Gaussian fit restricted to the same frequency and width results in an integrated flux density of $0.17 \pm 0.2$ Jy km s$^{-1}$, indicating no evidence for a line at that location.

 Both \cone\ and \ctwo\ were also contained within the field-of-view of an additional ALMA Band 3 program, 2015.1.00861.S.  Both sources were again far out in the primary beam of these data, at approximately the 0.25 and 0.1 response points, respectively. These data are described in more detail in \citet{Silverman2018}, and have non-overlapping frequency coverage with our own data. We downloaded and imaged these data following the same procedure as for our own data. The new images reach a continuum sensitivity at phase center of 10$\mu$Jy/beam at 93.5GHz and a line sensitivity of 120-170$\mu$Jy/beam per 100 km s$^{-1}$ channel, approximately a factor of two higher than in our data. Neither source is detected in continuum in these data, and were not expected to be detected given the sensitivity, effective frequency, and position of our sources within the ALMA primary beam. We additionally searched these data for blindly-detected CO lines as in our own data, but found no significant emission lines. The limited (primary beam-corrected) sensitivity of these data preclude us from drawing strong conclusions about the redshifts of either source.

For convenience, we define an equivalent survey area as the total area across all five ALMA maps at which a source with the same $S_{\footnotesize{\textrm{3mm}}}$ as \cone\ would be detected at $>5\sigma$.  Taking into account the small variations in the central frequency of each map (tuned to the specific redshift of the target $z\sim$1.5 galaxies), which change the primary beam response shape and the detection threshold (25-31$\mu$Jy) due to the steep spectral index of dust emission, we derive a total survey area of 8.0 arcmin$^{2}$ to 155$\mu$Jy ($5\sigma$ limit).

\subsection{Multi-wavelength photometry}

\cone\ has extremely deep coverage at all optical-to-sub-mm wavelengths, yet it has no counterpart within a radius of 3.3 arcseconds in the deepest published multiwavelength catalogs in COSMOS to date, from $0.6-1100\mu$m \citep{Laigle2016, Muzzin2013, Aretxaga2011, Geach2017, Hurley2017,LeFloch2009} or  $3-1.4$ GHz \citep{Schinnerer2010,Smolcic2017}. 
The astrometry of multi-wavelength catalogs in COSMOS are excellent owing to the registration between VLA observations \citep[e.g.][]{Schinnerer2007}, ground-based optical and space-based facilities, with an astrometric accuracy of 5 milli-arcseconds \citep{Koekemoer2007}. Similar astrometric accuracy was found between ALMA and COSMOS multiwavelength datasets \citep{Schreiber2017}.
There is no apparent flux at optical-to-Spitzer/IRAC wavelengths at the ALMA position (Figure \ref{cutout}).
 \ctwo\ is also missing from these catalogs, likely because it is blended with a bright neighboring galaxy $\sim1\farcs3$ to its north with photometric redshift 0.95 (see Figure \ref{cutout}), but appears detected in $Ks$ and IRAC. It is possible that the ALMA source is simply a highly obscured region in the low-redshift galaxy, which can be ruled out by spatially deblended SED analysis. We therefore proceeded to perform deblended photometry on both ALMA positions using the following data sets.

\subsubsection{Optical, near-infrared, and Spitzer/IRAC}

The optical data consists of the Subaru/Suprime-Cam  $B_j$, $V_j$, $g^+$, $r^+$, $i^+$ and $z^+$-imaging~\citep{Taniguchi2007}, with $5\sigma$ limits of $\sim25-27.4$\,mag in $1\farcs2$ apertures, and Subaru HyperSuprimeCam (HSC) $g$, $r$, $i$, $z$ and $y$ (\citealt{Aihara2018a, Aihara2018b}) imaging ($\sim25-26.8$\,mag, $5\sigma$). Ultradeep NIR coverage is provided by the 4th data release of the UltraVISTA survey ~\citep{Mccracken2012}, thanks to mosaics in the $Y$, $J$, $H$ and $K_s$ filters to $\sim25$\,mag (AB, $5\sigma$). Remarkably, the coverage in the $K_\mathrm{s}$ band from DR4 is $\sim0.9$\,mag deeper than from DR3, allowing us to place strong constraints on the flux density of \cone\ at NIR wavelengths. Stacked images were constructed using the optical imaging $0.4-0.8$ micron and in the near-infrared imaging $0.9-1.6$ micron.
We use \textit{Spitzer}/IRAC $3.6\mu$m and $4.5\mu$m mosaics that combine data from the S-COSMOS (\citealt{Sanders2007}) and the {\it Spitzer} Large Area Survey with HSC (SPLASH, PI: Capak) programs ($\sim24.5$\,mag, $5\sigma$ in $1\farcs8$ apertures), and the $5.8\mu$m and $8.0\mu$m from the S-COSMOS program ($\sim20.7$\,mag, $5\sigma$ in $1\farcs8$ apertures).

We measured flux densities for \cone\ in the optical and UltraVISTA bands in $1\farcs2$-diameter apertures after subtracting the neighbors using \textsc{mophongo} (\citealt{Labbe2013, Labbe2015}). This procedure carefully models the light profiles of the sources using a higher resolution image as a prior, minimizing potential contamination by bright nearby objects (see Figure \ref{deblend}). In our analysis we adopted the HSC $z$-band image as prior for \cone\ as it provides the best compromise between depth and resolution and the F814W-band image for \ctwo, given the nearby bright neighbor at $\sim1\farcs3$ distance.  Because of the broader PSF, we adopted $1\farcs8$ apertures for our estimates in the IRAC bands. Total flux densities were then estimated from the spatial profile and the relevant PSF-correction kernel.

\subsubsection{Far-infrared to submillimeter} 

Far-infrared and sub-millimeter {\it Herschel} fluxes 
were measured for both galaxies by simultaneously fitting Gaussian profiles at fixed prior locations in the {\it Herschel} images specified by the ALMA locations and augmented by the MIPS positions of all neighboring objects from the S-COSMOS data \citep[][]{Sanders2007,LeFloch2009}. 
The FWHM of the Gaussians were 7.7, 12" (PACS 100, 160$\mu$m) and 18, 25 and 37" (SPIRE 250, 350, 500$\mu$m), respectively. Uncertainties were computed by fitting gaussians at random locations within a 2 arcmin radius and computing the rms. Flux calibration was performed by comparing to the 24$\mu$m prior catalog of {\it Herschel} DR4 \citep[][]{Oliver2012, Hurley2017}. Comparison of the measured SPIRE fluxes with those published 
showed a scatter of 30\%; this was added in quadrature to the flux uncertainties for those bands. The correction is minor considering the signal to noise ratio of $\lesssim1$ of the SPIRE fluxes. For SPIRE 500$\mu$m only, the flux of a neighboring $z_{spec}=1.45$ galaxy 5'' to the south was subtracted separately prior to deblending, using its predicted IR emission based on the infrared SED of \citet{Wuyts2011} and its SFR$_{24,IR}$=65 M$_\odot$/yr.

The procedure for measuring photometry at 850$\mu$m follows that of {\it Herschel}/PACS and SPIRE (250 and 350$\mu$m) using the ALMA and MIPS positions as priors. The sources lie in a region of shallower ($\sigma_{850}\sim$3 mJy/beam) coverage in the inhomogenous SCUBA-2 Cosmology Legacy Survey \citep[S2CLS;][]{Geach2017} and are undetected in this map. We therefore use the state-of-the-art S2COSMOS observations (Simpson priv. comm.), which achieved a $\sigma_{850}\sim1.2$mJy over the entire field (Simpson et al. in prep).

\subsubsection{Radio VLA 3Ghz and 1.4 Ghz}
Neither source has a counterpart in either the VLA 3 GHz 5$\sigma$ source catalog \citep{Smolcic2017} or the 1.4 GHz deep survey 5$\sigma$ source catalog \citep{Schinnerer2010}.  Photometry on the 3GHz map using the ALMA position as prior reveals 9.98$\pm$2.39 $\mu$Jy/beam (4$\sigma$) point source for \cone, below the detection limit of the \citet{Smolcic2017} catalog. Blindly detected sources at this flux density have a high probability ($>$50\%) of being spurious \citep{Smolcic2017}, therefore requiring the ALMA prior in order to be considered real. \ctwo\ shows no significant flux to 3$\sigma$ limits of $\sim7\mu$Jy/beam. Within 10 arcmin$^2$, the 1.4 GHz map has rms $\sim$17 $\mu$Jy/beam and no significant flux ($>3\sigma$) is detected from either galaxy. 

\subsection{Spectral Energy Distribution Modeling}
\label{sec:sed}
The deep photometry from $\lambda=0.4-3000\mu$m places strong constraints on the SEDs and allows us to model their stellar populations and redshift. It should be noted that even if a source is not formally detected ($>3\sigma$) at most wavelengths, absence of flux can still provide useful constraints, in particular on the redshift. All fitting is performed in linear fluxes and uncertainties and no upper limits are enforced on measurements with low SNR.

We use the Bayesian Analysis of Galaxies for Physical Inference and Parameter EStimation (\textsc{bagpipes}) code \citep{Carnall2018}, which assumes the stellar population synthesis models of \citet{Bruzual2003} and implements nebular emission lines following the methodology of \citep{Byler2017} using the \textsc{cloudy} photoionization code \citep{Ferland2017}. We select the flexible dust absorption model of \citet{Charlot2000}, with the exponent of the effective absorption $\propto\lambda^{-n}$ as a free parameter and adopt the \citet{DraineLi2007} dust emission model under the assumption of energy balance, such that dust-absorbed light is re-radiated in the far-infrared. All stellar populations have this effective absorption, while the youngest stars (defined as those with age $<10$Myr) have an extra factor of attenuation applied ($\eta$) to account for dusty birthclouds.

The dust emission model is parameterized using the starlight intensity $U$ incident on the dust (translating into a distribution of dust temperatures), the amount of PAH emission, and the fraction of dust at the coldest temperature. We further assume a delayed exponential star-formation history with $\tau$ and age left free and metallicities ranging from $0.2-2.5 Z_\odot$. For each parameter, uniform, diffuse priors are assumed (see Table \ref{tab:sedfit}). In general, we do not expect to constrain most free parameters, but we will marginalize over those and limit the discussion to the more important parameters including redshift, stellar mass, and SFR. Where relevant, we will also quote results based on assumptions often used in the literature, such as assuming a fixed \citet{Calzetti2000} dust attenuation model.

\begin{figure*}[th]
    \centering
    \includegraphics[scale=.5, trim=70 0 0 0,clip]{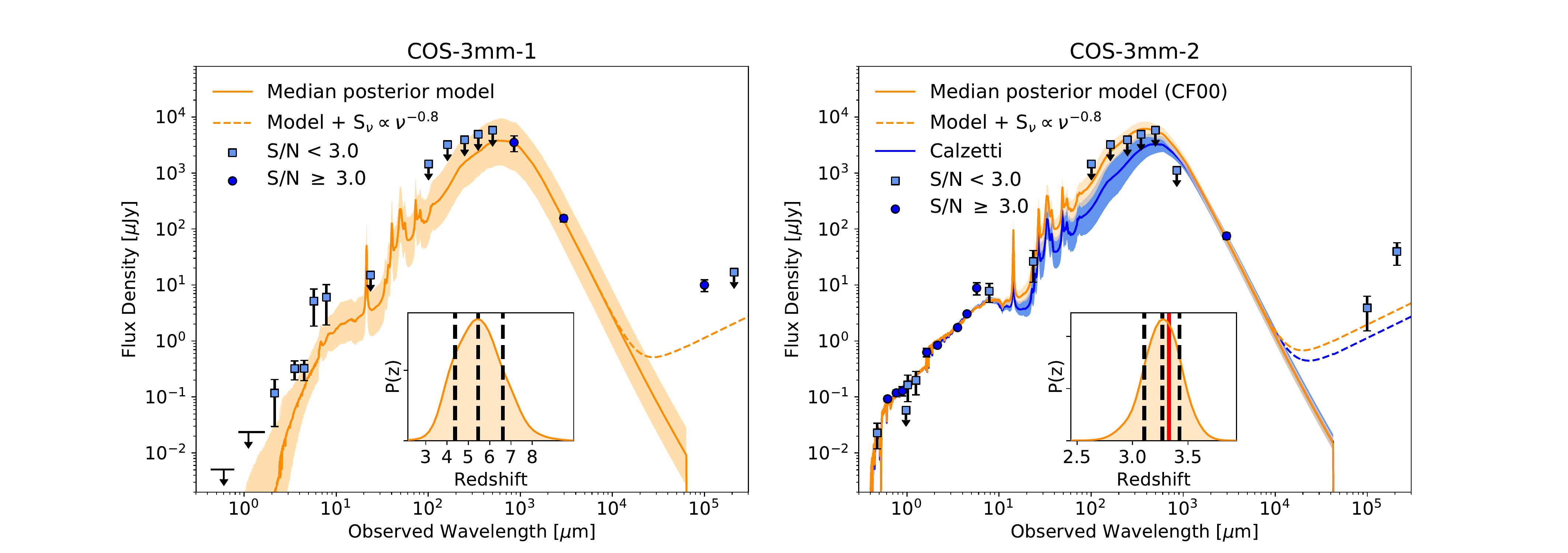}
    \caption{ Left: Observed photometry of \cone\  (points). For display purposes, we show   1$\sigma$ upper limits (downward arrows) 
     for data with $SNR<1$ and stacked optical and NIR fluxes (horizontal bars) instead of the individual optical/near-IR non-detections. Any photometry with $SNR>1$ is plotted with $1\sigma$ uncertainties but does not necessarily indicate a significant detection.
      Light blue squares indicate photometry with SNR$<$3, and dark blue circles indicate detections with SNR$\geq$3. Shown are the median posterior spectrum (dark orange) and 16-84th percentile range (light orange) from Bayesian SED fitting with \textsc{bagpipes} assuming the \citet{Charlot2000} dust model. Insets contain the posterior redshift distribution and the 16, 50, and 84th percentiles (black lines).
    The deep FIR photometric limits at $24<\lambda<250\mu$m favor high redshift (90\% probability at $z>4.1$). Dotted line indicates the radio spectrum $\propto \nu^{-0.8}$ expected from the total L$_{\textrm{IR}}$ \citep{Tisanic2019}, suggesting a 3 Ghz radio excess.  Right: SED of \ctwo. As in the left panel, any photometry with $SNR>1$ are plotted as errorbars with $1\sigma$ uncertainties. The photometric redshift agrees with CO($4-3$) at $z_{CO}$=3.329 (red line). 
The SED fit (orange) assumes the \citet{Charlot2000} dust model and the redshift fixed to $z_{CO}$. Also shown is a fit assuming a \citet[][]{Calzetti2000} starburst attenuation law (blue curve) a common assumption in high-redshift studies, which can drastically change the estimated stellar mass, LIR, and SFR. The blue curve has $8\times$ lower stellar mass and 2$\times$ lower SFR, while still adequately fitting the observations.     }
    \label{sed}
\end{figure*}

\section{Results}

The photometry for both galaxies is presented in the Appendix (Table \ref{tab:phot}).

\subsection{\cone}

\cone\ is undetected ($<3\sigma$) in any individual optical ($>27$AB), near-infrared ($>25.2$), and Spitzer/IRAC ($>25.2$) bands, and remains undetected in the stacked ultradeep optical ($>29.6$ AB) and near-IR ($>27.9$ AB) data ($1\sigma$ limits). The source is faintly detected (25.2 AB $\sim3\sigma$) in deep stacked IRAC 3.6+4.5 observations, indicating the source is likely real with extremely red colors. Owing to the shallower depth, the source is undetected in deep Spitzer/MIPS 24 micron and {\it Herschel} PACS+SPIRE $160-500$ micron. The flux measured with SCUBA is $S_{850}$=3.5$\pm$1.1 mJy ($SNR=3.1$). 
 This flux density is fainter than the depth typically achieved by deep single dish sub-mm surveys \citep[$>$3.5-5 mJy; e.g.][]{Weiss2009,Aretxaga2011,Geach2017,Cowie2017} at robust detection thresholds ($\gtrsim4\sigma$). 
The observed SED is shown in Figure \ref{sed} (left panel). For clarity, photometry with S/N $<$ 1 are indicated with arrows at the 1$\sigma$ rms level, and S/N $>$ 1 with 1$\sigma$ errorbars. The deep non-detections at all wavelengths between $24-500\mu$m and the extreme flux ratios 
($S_{24}/S_{850}<4\times10^{-3}$, 
$S_{24}/S_{3mm}<9\times10^{-2}$, 
$S_{4.5}/S_{850}<9\times10^{-5}$, 
$S_{4.5}/S_{3mm}<2\times10^{-3}$, see e.g., \citealt{Cowie2018,Yamaguchi2019}), demand that dust peak is highly redshifted $z>4$.

Fitting stellar population models, we find the observations are well reproduced by a massive $10^{10.8\pm0.4}$M$_\odot$, star forming  $SFR=309^{+241}_{-149}$M$_\odot$/yr,  highly obscured $A_V\sim4^{+1.4}_{-1.0}$ galaxy at very high redshift $z\sim5.5^{+1.2}_{-1.1}$. The SED fitting results are shown in Figure \ref{sed} and the posterior values and priors are listed in Table \ref{tab:sedfit}. The posterior distributions of all parameters are presented in Figure \ref{fig:pdf1}. 
Adopting a classical Calzetti dust law would increase the stellar mass, SFR, and redshift by 60\%, 10\%, and 5\% respectively. These changes are within the uncertainties, so we elect to adopt our more conservative values measured assuming \citet{Charlot2000} as fiducial. We also experimented with fitting only the mid-infrared to sub-millimeter SED, finding the same results. In all cases, the solutions require the galaxy to be at high redshift, with  $z>4.1$ at 90\% confidence {and a total infrared luminosity $L_{IR}=4\times10^{12}L_\odot$ (based on integrating the median posterior spectrum). } Adopting an SED typical for the most obscured $A_V>3.5$ SMGs \citep{daCunha2015} produces similar redshift and LIR, as does adopting the Arp 220 SED. Converting to SFR using the \citep{Kennicutt2012} conversion, which does not implicitly assume a SFH, results in a larger estimated SFR$=540$M$_\odot$/yr, but consistent within the 1$\sigma$ uncertainties from SED fitting.

\begin{figure*}[th]
    \centering
    \includegraphics[scale=.5]{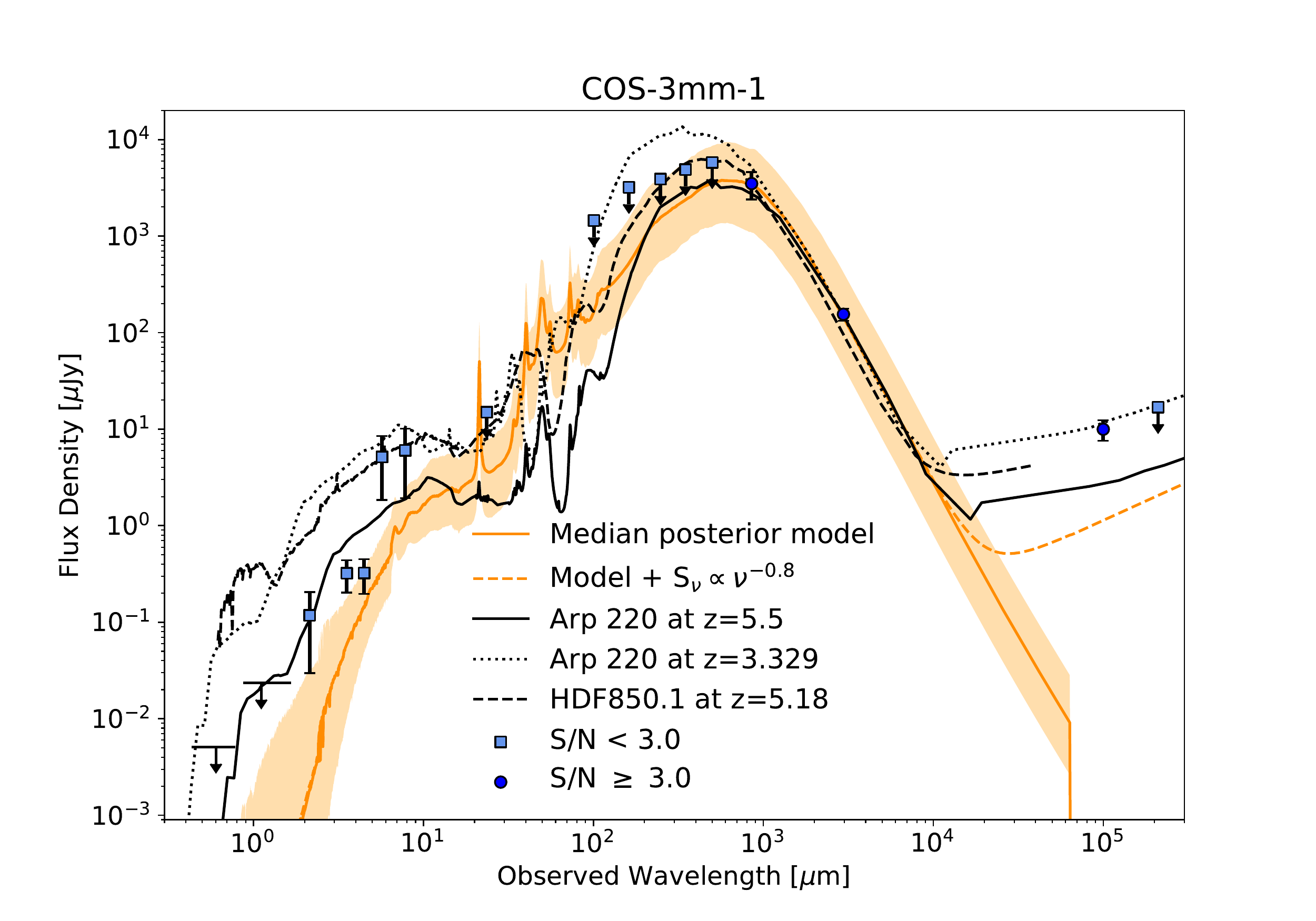}
    \caption{Comparison between the SED of \cone\ with that of both HDF850.1 and Arp220. SED fitting results for \cone\ as in the left panel of Figure \ref{sed}. The SED of Arp220 scaled to the 850$\mu$m and 3-mm photometry of \cone, at the photometric redshift of \cone, is remarkably similar to the \cone\ SED. Forcing Arp220 to the redshift of \ctwo\ violates the Herschel FIR+sub-mm constraints.
    The SED of \cone\ is very similar to that of the bright z$_{spec}$=5.18 dusty galaxy HDF850.1 (using the SWIRE SED of IRAS 20551-4250 as in \citealt{Serjeant2014}) scaled to match the average observed 850$\mu$m and 3-mm flux).  }\label{fig:hdf850}
\end{figure*}

\begin{deluxetable*}{cccccc}[t!]
\tablecaption{Stellar Population Properties of 3-mm continuum sources} 
\tablecolumns{5}
\tablewidth{0pt}
\tablehead{
\colhead{Parameter} &
\colhead{\cone$^a$} &
\colhead{\ctwo} &
\colhead{\ctwo} &
\colhead{Range (uniform prior)} \\
\colhead{Attenuation Curve} & 
\colhead{CF00} & \colhead{CF00} & \colhead{Calzetti}
}
\startdata
Coordinates & 10:02:36.82 +02:08:40.60  & 10:02:36.30 +2:08:49.55 & \\
Redshift & 5.5  $^{+ 1.2 }_{- 1.1 }$ &  z$_{CO}$=3.329 & z$_{CO}$=3.329 & (0.0, 10.0) \\ 
SFR [M$_{\odot}$ yr$^{-1}$]  & 309  $^{+ 241 }_{- 149 }$ &  247  $^{+ 77 }_{- 76 }$ &  112  $^{+ 27 }_{- 38 }$ &  \\ 
Log$_{10}$ Stellar Mass [M$_{\odot}$]  & 10.8  $^{+ 0.4 }_{- 0.4 }$ &  11.1  $^{+ 0.2 }_{- 0.2 }$ &  10.2  $^{+ 0.2 }_{- 0.2 }$ &  (6.0, 14.0) \\ 
Mass-weighted age [Gyr]& 0.2  $^{+ 0.1 }_{- 0.1 }$ &  0.5  $^{+ 0.1 }_{- 0.2 }$ &  0.1  $^{+ 0.2 }_{- 0.1 }$ & \\ 
A$_V$ & 4.0  $^{+ 1.4 }_{- 1.0 }$ &  2.7  $^{+ 0.3 }_{- 0.3 }$ &  1.4  $^{+ 0.1 }_{- 0.1 }$ &  (0.0, 10.0) \\ 
$\eta^c$ & 2.0  $^{+ 0.6 }_{- 0.6 }$ &  2.1  $^{+ 0.6 }_{- 0.8 }$ &  2.5  $^{+ 0.3 }_{- 0.5 }$ &  (1, 3.0) \\ 
Umin$^d$ & 13.0  $^{+ 7.7 }_{- 7.3 }$ &  7.6  $^{+ 6.3 }_{- 4.0 }$ &  4.8  $^{+ 3.4 }_{- 2.4 }$ &  (1, 25.0) \\ 
gamma$^e$ & 0.5  $^{+ 0.3 }_{- 0.3 }$ &  0.5  $^{+ 0.3 }_{- 0.3 }$ &  0.5  $^{+ 0.3 }_{- 0.3 }$ &  (0.01, 0.99) \\ 
n$^f$ & 0.9  $^{+ 0.6 }_{- 0.4 }$ &  0.4  $^{+ 0.1 }_{- 0.1 }$ &  - & (0.1, 2.0) \\ 
qpah$^g$ & 2.3  $^{+ 1.1 }_{- 1.1 }$ &  3.3  $^{+ 0.5 }_{- 1.0 }$ &  3.0  $^{+ 0.7 }_{- 1.1 }$ &  (0.5, 4.0) \\ 
Log$_{10}$ LIR$^h$ [L$_{\odot}$]  & 12.6 & 12.4 & 12.1 \\ 
Gas mass [M$_{\odot}$]  & 0.5-1.5 $\times 10^{11}$ & $4-9\times10^{10}$ & \\
\enddata
\tablenotetext{a}{Fitting priors are uniform with the range as defined in last column.}
\tablenotetext{b}{Redshift prior is a gaussian matching the redshift evolution at $z>4$ cumulative halo mass function as described in Section \ref{sec:sed}}
\tablenotetext{c}{Multiplicative factor producing extra attenuation for young stars}
\tablenotetext{d}{Starlight intensity on dust grains, related to dust temperature as T$_{dust}\propto U^{1/6}$  \citep{DraineLi2007}   }
\tablenotetext{e}{Fraction of stars at Umin}
\tablenotetext{f}{Power-law slope of the dust extinction law (for \citealt{Charlot2000})}
\tablenotetext{g}{PAH mass fraction}
\tablenotetext{h}{Total infrared luminosity (8-1000$\mu$m) in units  from integrating the median posterior spectrum in Figure \ref{sed}.}
\label{tab:sedfit}
\end{deluxetable*}

Calculating redshift using the radio-to-submm spectral index method (using 850$\mu$m and the upper limit on 1.4GHz) implies a lower limit to the photometric redshift of $\sim$4.7 \citep[e.g.][]{CarilliYun1999}. We note that this is very similar to the massive dusty galaxy HDF850.1 which had a redshift 4.1 according to this relation \citep{Dunlop2004}, and was later spectroscopically confirmed at z=5.2 \citep{Walter2012}. The SED of \cone\ is very similar to that of HDF850.1; scaling its best-fit SED to the 850$\mu$m and 3-mm fluxes of \cone\ predicts its observed 3 GHz radio flux of 10$\mu$m (Figure \ref{fig:hdf850}). 

The observed 3 GHz radio flux ($4\sigma$) is in excess of that expected from empirical SED templates for obscured star-forming galaxies \citep[e.g. SMGs;][]{daCunha2015} and recent calibrations of the redshift-dependent ratio of total infrared luminosity to rest-frame  1.4 GHz luminosity density \citep[][]{Delhaize2017, Tisanic2019}. For typical assumptions of the radio spectral slope $S_{\nu}\propto\nu^{-0.8}$ (dashed line in Figure \ref{sed}), the radio emission of \cone\ is a factor of $\sim6\pm 1.5$ higher than expectations for star-formation, in excess of the commonly adopted $\times$3 threshold for AGN activity \citep{Daddi2009}.  No evidence for redshift evolution in the spectral slope, or dependence on sSFR or distance from the main sequence has been reported \citep{Magnelli2015}.

The radio excess therefore indicates plausible evidence for presence of an AGN. We note that no X-ray counterpart exists within 5'' \citep{Civano2016}, although this is not surprising given the high-redshift of \cone. 
Given the possible presence of an AGN, it is worth noting that inferred parameters from SED fitting could be biased by AGN emission, in particular in the mid-infrared \citep[e.g.][]{Leja2018}. Therefore, we investigate any possible contamination using the empirical AGN-dominated template published by \citep{Kirkpatrick2015}. We use their mid-infrared template based on galaxies with the largest mid-infrared luminosity contribution from AGN emission scaled to the 24$\mu$m flux of \cone\, and subtract the predicted AGN contribution from the observed SED. This represents the maximum AGN contribution that could be accommodated by the data without violating the observed photometry. We then re-measure the SED fitted parameters. Subtracting this empirical AGN template reduces the SFR by $\sim$15\%, well within the uncertainty, but does not change the inferred stellar mass significantly. We conclude that any AGN contribution does not significantly impact our SED-fitting results.

Finally, the 3-mm flux density enables an estimate of the molecular gas mass for \cone\ following the calibration of sub-mm flux to gas mass \citep{Scoville2016}. Exploring the range of the photometric redshift PDF, and typical assumptions about the dust emissivity, temperature, and dust-to-gas ratio, we find that \cone\ likely has M$_{\rm{gas}}\sim0.5-1.5 \times 10^{11}$ M$_{\odot}$. This is independent evidence that the galaxy is massive, with a high inferred gas fraction ($\sim$60\%).

\subsection{\ctwo}

\ctwo\ is optically faint, but detected in $K_{s,AB}=24.2$ and relatively bright in IRAC $\sim23$ mag, with red optical-to-IRAC colors. There is an apparent break between the $J_s$ and $H$-band, consistent with a Balmer/4000\AA\ break at $z\sim3$. The source is undetected at $24-870$ micron, but with less extreme flux ratios ($S_{24}/S_{3mm}<3.5\times10^{-1}$, $S_{4.5}/S_{3mm}<3\times10^{-2}$)  compared to \cone, indicating a lower redshift.

We follow the same procedure as before to fit the SED, finding a well-constrained photometric redshift of $z=3.3\pm0.2$ (99\% probability between $2.7 < z < 3.7$), ruling out that this source is simply a highly obscured region in the nearby $z=0.95$ foreground galaxy to the north. Considering the CO emission line detection (consistent with redshifts 2.25, 3.33, and 4.41), we determine the line is likely CO($4-3$) at $z=3.329$, in agreement with the best-fit photometric redshift. Fixing the redshift to the spectroscopic redshift, the observations are best reproduced with a massive $10^{11.1}$M$_\odot$, star forming $SFR=250$ M$_\odot$/yr, highly obscured $A_V\sim2.7$ galaxy. Using the \citet{Kennicutt1998,Kennicutt2012} conversion from LIR to SFR yields $\sim375$ M$_{\odot}$ yr$^{-1}$. {The derived gas mass using the CO(4-3) line flux (assuming the average CO excitation from \citealt{Bothwell2013}) is in the range $\sim1-4\times10^{11}$ M$_{\odot}$, given the factor of four scatter in measured excitation for high-redshift dust obscured galaxies \citep[e.g. Figure 1 in][]{Narayanan2014}. The 3 mm derived gas mass is $4-9\times10^{10}$ M$_{\odot}$. Both estimates indicate gas fractions in the range $30-70$\%.} The SED properties for \ctwo\ are listed in Table \ref{tab:sedfit}, the SED fit is shown in Figure \ref{sed}, and posterior distributions for the parameters are presented in Figure \ref{fig:pdf2}.

The best-fit power law index for the \citet{Charlot2000} dust model is $n=0.4\pm0.1$, flatter than the $n=0.7$ appropriate for typical nearby starburst galaxies. The flat spectral index means that the attenuation curve is ``greyer'' than a Calzetti attenuation law \citep[e.g.][]{Chevallard2013} resulting in larger attenuation for the same amount of reddening $E(B-V)=0.3$. The assumed attenuation model can have a large impact on the derived stellar mass. If instead of a flexible attenuation model a classical Calzetti starburst law is assumed then the fits result in a factor $8^{+7.1}_{-3.9}$  lower stellar mass ($10^{10.2}$ M$_\odot$) and factor $2.2\pm0.7\times$ lower SFR (same answer if the infrared-to-mm constraints are also removed). The large difference in stellar mass is remarkable. The reason is that the steeper reddening curve of Calzetti reproduces the data with an intrinsically blue, young, low M/L ratio stellar population, driving the lower masses. The \citet{Charlot2000} model requires an older, redder, higher M/L population, while also implying larger attenuation at all wavelengths, including the near-infrared. We note that these results all assume energy balance and a parametric delayed$-\tau$ SFH. We explore the dependence on SFH by also fitting the SED with \textsc{prospector} \citep{Leja2017}, which is capable of fitting a non-parametric SFH in logarithmic bins of age \citep{Leja2019}. These fits produce qualitatively similar results, with Calzetti yielding $\sim2\times$ lower SFR and stellar mass.

An independent estimate of mass can be derived by estimating the dynamical mass from the 630 km/s width of the CO emission line at 106.5 GHz (Figure \ref{fig:CO43}). Following the procedure used by \citet{Wang2013}, we compute $V_{circ} = 0.75  FWHM(CO)/sin(i)$ and $M_{dyn}=1.16 \times 10^{5} V_{circ}^2 D$, where we note that inclination angle $i$ and disk diameter D (in kpc) are unknown. Adopting the mean size $R_e=3^{+1.8}_{-0.9}$ kpc expected for massive 10$^{11}$M$_\odot$ star forming galaxies at $z=3$ \citep{vanderWel2014}, we find M$_{dyn}sin^2(i) = 1.5^{+1.0}_{-0.5}\times10^{11}$M$_\odot$. Assuming a disk geometry with an inclination angle of $i=45^{\circ}$ would imply M$_{dyn}\sim3\times10^{11}$M$_\odot$. Overall, the high dynamical mass appears to agree better with the high stellar mass inferred from \citep{Charlot2000} dust model with flat $n$ than with the Calzetti-based stellar mass, but the large uncertainties in dynamical and stellar mass prevents firmer conclusions.
 
\subsection{Redshifts}
Given the close $r\sim15''$ separation on the sky of the two ALMA sources we consider the possibility that they are at the same redshift. The redshift of \ctwo\ ($z=3.329$) is strongly disfavored for \cone\  (1\% probability $z<3.3$). Alternatively, our identification of $z_{CO(4-3)}=3.329$ for \ctwo\ is erroneous and the galaxy is at $z_{CO(5-4)}=4.41$, more consistent with \cone. This is unlikely based on the SED fit of \ctwo, which is well constrained by the presence of both a Lyman Break and Balmer break.
We find no significant emission line ($SNR>3$) in the spectrum of \cone\ which suggests it is unlikely at $3.20 < z < 3.35$ and  $3.72 < z < 3.90$.
A Gaussian fit to the spectrum of \cone\ at fixed $106.5$ GHz indicates $SNR=0.8$, providing no evidence for \cone\ and \ctwo\ occupying the same dark matter halo (DMH). Additionally, we scan the nearby velocity space in case both galaxies are in a large scale structure filament, but we find no emission line $SNR>2$ within $\Delta v=\pm2000 $km/s. Finally, we consider the odds of finding two 3-mm sources within $r\sim15''$ on the sky. Using the 3-mm number counts of \citet{Zavala2018_3mm}, we simulate uniformly random distributions of two sources which indicates an $\sim10$\% chance of finding a $75\mu$Jy source within $r<15''$ of a $155\mu$Jy source. These odds are low, but not negligible. Overall, there is no conclusive evidence that the two sources are at the same redshift. We therefore proceed and take the $z_{phot}=5.5\pm1.1$ for \cone\ at face value.

\section{Discussion}

\subsection{Implications from full optical-to-mm spectrum SED fitting} 

We consider the SED fitting analysis using a stellar population model with self-consistent dust absorption and emission under the assumption of energy balance and constrained by high quality optical-to-millimeter observations. In the case of the $z\sim5.5$ source \cone, which lacks strong detections at any wavelength other than 3 mm, it is remarkable that the Bayesian posterior probability distributions for key parameters such as redshift, stellar mass, dust attenuation and star-formation rate are as constrained as they are
(see Table \ref{tab:sedfit} and Figure \ref{fig:pdf1}). Closer inspection indicates that the results are mostly driven by the combination of the high SNR ALMA measurement with deep photometric limits in the short wavelength infrared ({\it Spitzer}/MIPS and {\it Herschel}/PACS $100-160\mu$m), which demand that the dust emission peak is highly redshifted ($z>4$) and the SFR is high. It is likely that the uncertainties on the stellar mass and other stellar population parameters are underestimated, because of the tight prior  imposed by the parametric SFH \citep{Carnall2019,Leja2019}. We note that our choice of a rising SFH produces a relatively conservative (lower) estimate of stellar mass compared to constant or declining SFH.

There is accumulating evidence that the attenuation law in very dusty galaxies can be flatter than Calzetti \citep{Salmon2016, Leja2017, Salim2018}, possibly caused by a more uniformly mixed geometry of both old and young stars with dust \citep{Narayanan2018}. The flat inferred attenuation law $\lambda^{-0.4}$ for \ctwo\ is not surprising in that regard. It is notable and sobering however that modeling the optical-to-millimeter with a classical Calzetti starburst law instead (keeping everything else the same) results in significantly lower SFR and stellar mass ($10^{10.2}$), in apparent tension with the high dynamical mass ($1.5-3\times10^{11}$M$_\odot$). Detailed analyses of ULIRGs at $z\sim2$ with photometric constraints on far-infrared dust emission \citep[e.g.][]{LoFaro2017} also found that stellar masses inferred using a Calzetti law are systematically lower because of the smaller amount of reddening at near-infrared wavelengths. Clearly, caution should be exercised when applying a single locally-calibrated attenuation law at high redshift. Overall, the results highlight the need for deep (sub-)mm  measurements to determine bolometric luminosities and provide high-redshift empirical constraints on the dust law/energy balance \citep[e.g.][]{Hodge2016}.

For the remainder of the paper we emphasize discussing the implications of the ALMA-only source \cone, which likely represents a population that is absent from all current optical-IRAC selected galaxy studies, is below the nominal detection threshold of deep single-dish sub-mm surveys, and therefore deserves more scrutiny.

\subsection{Number counts}\label{sec:ncounts}
A recent unbiased ALMA 3 mm archival search for continuum sources over 130 independent pointings
\citep{Zavala2018_3mm} enabled the first 3 mm number counts. For archival searches, however, detection limits and effective search area are necessarily very inhomogeneous due to the strong variation of the ALMA primary beam response, complicating a straightforward comparison of results. Restricting the comparison to 
\cone, we use their best-fit powerlaw to the cumulative number counts, which considers effective selection area and incompleteness, predicting one source brighter than S$_{3\textrm{mm}}>155\mu$Jy per $16^{+16}_{-8}$ arcmin$^2$. This is consistent with our finding one source in our effective survey area of 8 arcmin$^2$ (see $\S$\ref{sec:almadet}).

We note that given the estimated stellar mass of $\sim 10^{10.8}$M$_\odot$ at z=5.5, we expect \cone\ to occupy a massive DMH ($\sim 10^{12}$M$_\odot$) and to be strongly clustered. While our results are dominated by Poissonian uncertainties, this may imply that the uncertainties in published number counts are underestimates. Clustering may affect source counts from studies over very small areas such as the ASPECS 4.6 arcmin$^{2}$ survey in the HUDF \citep{GonzalezLopez2019}, and even the counts in the multiple pointings of \citet{Zavala2018_3mm}, as these observations were generally targeting moderately-sized deep fields (COSMOS, CDFS, UDS). Clustering effects may be exacerbated by the fact that dust obscuration preferentially occurs in massive galaxies \citep{Whitaker2017}. 
We note that our results are unlikely to be biased by the original primary targets, which were all at low redshift ($z\sim1.5$). Clearly, larger blind surveys are needed to constrain number densities at the bright end. 

\subsection{Unbiased ALMA 3 mm selection of \\ high-redshift galaxies}\label{sec:3mmsel}

Simulations of dust obscured galaxies in the early universe predict $3$ mm continuum selection optimizes the sensitivity to DSFGs at redshift $z>4$ \citep[][]{Casey2018b}. Current evidence is still mixed: continuum detected faint galaxies in the small ultradeep ALMA ASPECS field are at average $\bar{z}=2.3$ \citep{GonzalezLopez2019}, while limited ground-based spectroscopic evidence for sources found in wider-area archival data indicates $\bar{z}=3.1$ \citep[][]{Zavala2018_3mm}. Note, however, that the luminosity of the sources and volume covered could impact the redshift selection function \citep{Strandet2016,Brisbin2017}, and follow up ground-based spectroscopy in the optical and near-IR is possibly biased against highly obscured sources at $z>3$, due to the faintness of Lyman-alpha, nebular optical lines and lack of wavelength coverage $>2.4$ micron. Overall, the redshifts of the two continuum sources in this study, a probable $z_{CO}=3.329$ (based on a single CO line and congruous $z_{phot}=3.25\pm0.15$) and $z_{phot}=5.5\pm1.1$ ($\bar{z}=4.4$),  are consistent with 3 mm favoring higher redshifts than the $\bar{z}=2.2$ typical for $870\mu$m-selected galaxies \citep{Simpson2014}.

A challenge in determining the selection function is the difficulty in identifying counterparts and determining redshifts. Neither of our two 3 mm sources had counterparts in previous deep optical-to-radio selected catalogs, raising some concern for analyses where this is a critical step  (e.g., photometric redshifts or spectroscopic follow up in optical/NIR). This is particularly problematic for single-dish sub-mm observations, due to the large beam size, but here we find it challenging even with the high spatial resolution $FWHM=2''$ and accurate astrometry offered by ALMA. In the case of \ctwo\ the source is very close to a bright foreground galaxy (1.3'' to the North), which was initially mistaken as the counterpart. The optical/NIR faintness and low resolution of the Spitzer/IRAC data caused it to be missing or blended in existing multiwavelength catalogs. In the case of \cone\ the combination of high redshift and high obscuration resulted in it being extremely faint and undetected at all optical-infrared wavelengths. It is therefore possible that the identification of the highest redshift galaxies in existing (sub-)mm selected samples is still incomplete. In addition, obtaining spectroscopic redshifts in the optical/NIR is unfeasible for the faintest sources. ALMA spectral scans targeting CO and [CII] are likely the only recourse until other facilities (e.g., JWST/NIRSpec, LMT Redshift Search Receiver) become available.

\begin{figure*}[t]
    \centering
    \includegraphics[width=0.99\textwidth]{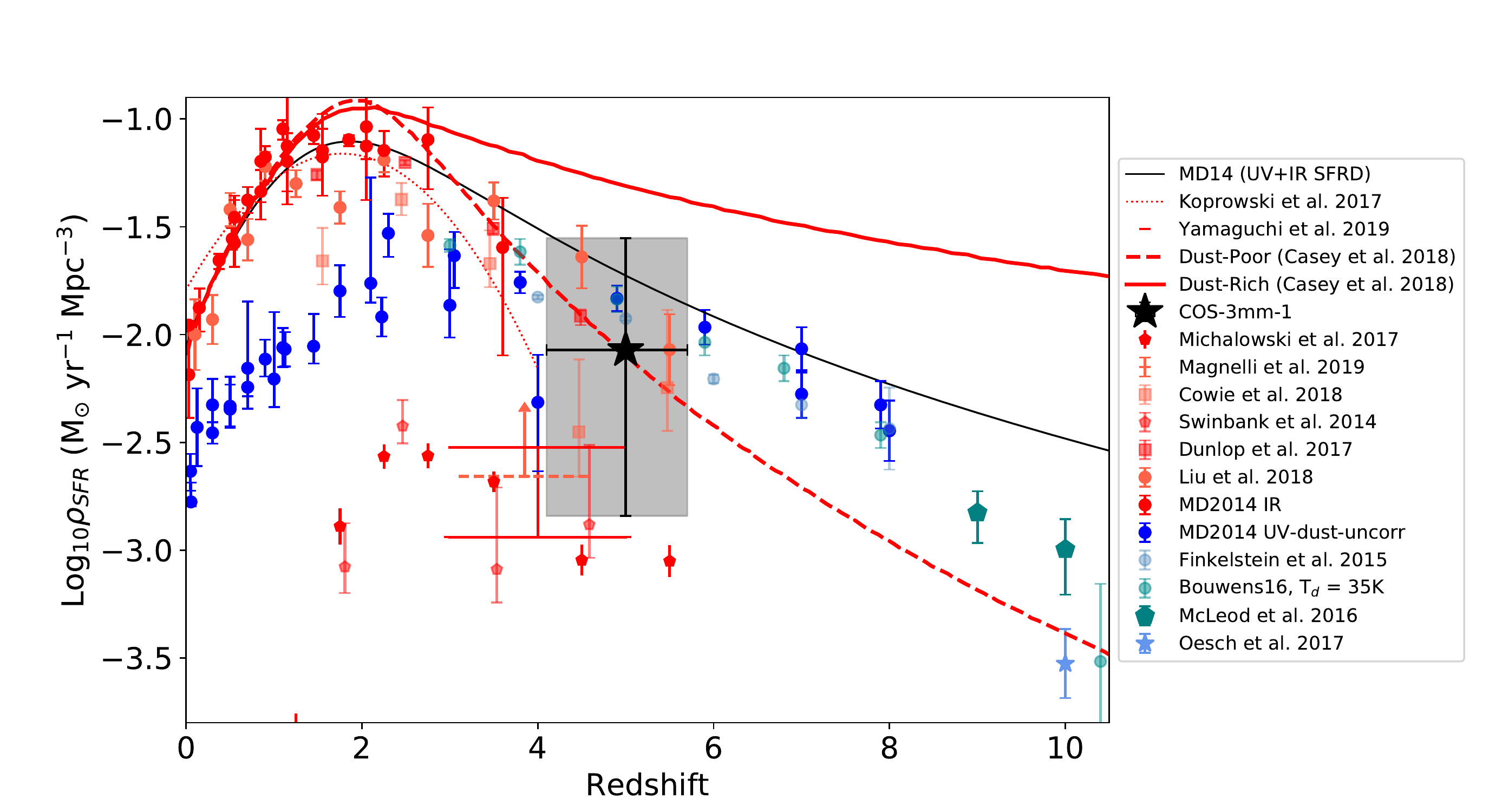}
    \caption{ The cosmic star-formation history of the Universe. Blue shades are UV (dust un-corrected) and red shades are IR-to-millimeter derived SFRs. 
    We add to the compilation of \citet[][; blue circles]{MadauDickinson2014} with more recent $z>4$ measurements by \citet{Finkelstein2015, Bouwens2016LF, McLeod2016, Oesch2018}. Red circles are the dust-obscured IR compilation of \citet{MadauDickinson2014}, to which we add recent $z>2$ measurements from \citet{Swinbank2014, Koprowski2017, Magnelli2019, Cowie2018, Dunlop2017, Liu2018}. 
     The contribution of \cone\ is indicated by the black star and shaded region, where the redshift range indicates the estimated selection volume as discussed in \S\ref{sec:sfrd}.      }
    \label{sfrd}
\end{figure*}

\subsection{Comparison to other ``dark" galaxies}

 The observed SED of \cone\ is different from that of other optical/near-IR/IRAC-faint populations.
The ALESS survey identified 9 IRAC-faint SMGs to 1 magnitude brighter IRAC limits \citep[][]{Simpson2014}, but these sources were much brighter in the far-infrared (peaking at $250-350$ micron to $\sim10$mJy), with a significantly bluer dust peak consistent with $z\sim2-3$. Such an infrared SED is ruled out by our observations. \cone\ is fainter at all wavelengths than the so-called H/K$-$dropout galaxies \citep[e.g.][]{Wang2016, Schreiber2017,Caputi2012}, which are generally selected on $H-4.5$ color, $>10\times$ brighter in IRAC, and estimated to be at lower redshift $\bar{z}=3.7$.

More recently, SCUBA surveys have been carried out to unprecedented depth over CANDELS fields 
\citep[$\sim1.5$ mJy detection limit; e.g.,][Simpson et al. in prep.]{Koprowski2017,Cowie2017,Cowie2018,zavala2017}.  Such surveys are deep enough to discover $S_{870}=2-4$ mJy objects like \cone\ at $z>4$. Indeed, the deepest SCUBA surveys have reported sub-mm selected objects that are likely at $z>4$ and that are either extremely faint at optical-IRAC and MIPS wavelengths \citep{Cowie2018} or likely initially misidentified and determined to be at high redshift based on the long wavelength IR SED \citep[e.g.,][]{Koprowski2016}. With optical-IRAC magnitudes in the range $29-25$ AB, no MIPS detection, and only a marginal 3 Ghz radio detection, \cone-like objects would have likely defied secure identification, and its probable $z=5-6$ redshift could have only been determined from the IR SED.

More similar are recently reported ALMA 1.2 mm selected sources from the ASAGAO survey \citep{Yamaguchi2019} in GOODS-S. These sources also lack obvious counterparts at shorter wavelength, show extremely small $S_{4.5}/S_{1.2mm}$ and $S_{24}/S_{850}$ flux ratios (indicative of high redshift), but are generally fainter $S_{1.2mm}=0.44-0.8$ mJy than \cone\ ($S_{1.2mm}\sim2.3$). Additional candidates that lack deep multi-wavelength ancillary data may exist outside legacy fields \citet{Wardlow2018}.

\subsection{Contribution to CSFRD}
\label{sec:sfrd}

The fact that \cone\ was identified in a survey of such small area suggests that similar galaxies are common in the early Universe. Using the effective area of 8 arcmin$^2$ as derived in \S\ref{sec:almadet}, our finding implies a source density of $0.13^{+0.30}_{-0.10}$ arcmin$^{-2}$, an order of magnitude higher than the rare sub-mm selected starbursts at $z>4$ ($0.01-0.02$ arcmin$^{-2}$; e.g. \citealt{Danielson2017, Marrone2018}).  If our results are representative, galaxies like \cone\ could represent the ``iceberg under the tip" of the known extreme dust-obscured star-forming galaxies in the early Universe. 

To estimate the space densities and contribution to the CSFRD requires estimating a selection volume, which is difficult as we do not know the expected properties or true abundance of DSFGs at $z>4$. Instead, we derive a simple estimate based on the derived properties of \cone. Encouraged by the strong redshift constraints from the long wavelength photometry, we set the lower bound of the selection volume to the lower 10th percentile redshift probability ($z>4.1$). The upper bound is chosen to be the redshift beyond which we expect only 0.1 halos of sufficient mass based on the cumulative halo mass function. 
Using the halo mass function calculator \textsc{HMF} published by \citet{Murray2013} and assuming the halo mass function of \citet{Behroozi2013} and a high $30\%$ baryon conversion into stars and (converting M$^*$=$10^{10.8}$M$_\odot$ into a conservative M$_{halo}$ $\sim10^{12}$M$_\odot$) we compute this upper bound to be $z=5.7$. 

Using this selection volume we determine a space density of $2.9^{+6.5}_{-2.4}\times10^{-5}$ Mpc$^{-3}$. We find that the contribution to the CSFRD by \cone\ is  $\rho_{\small \textrm{SFR}}$ $0.9^{+2.0}_{-0.7}\times10^{-2}$ M$_{\odot}$ yr$^{-1}$ Mpc$^{-3}$ (converted to a \citealt{Chabrier2003} IMF for comparison with literature measurements). Assuming instead the cumulative space density of massive halos ($>$10$^{12}$) at $z\sim5$ using the \citet{Behroozi2013} halo function, produces a very similar  $\rho_{\small\textrm{SFR}}$ $0.6\times10^{-2}$ M$_{\odot}$ yr$^{-1}$ Mpc$^{-3}$ for our derived SFR.

With only a single object, the Poissonian uncertainty is large and dominant \citep{Gehrels1986}. Cosmic variance is only on the order of $\sim$30\% based on the calculator by \citet[][]{Trenti2008}, and is therefore not further included. No completeness corrections are applied, because the true distribution is unknown. Figure \ref{sfrd} shows comparisons to various results at $0<z<10$ from literature that report dust-uncorrected UV-derived SFR and the dust-obscured SFR (IR to millimeter derived).

The contribution of \cone\ is  higher than inferred for the two near-infrared dark ALMA 1.2 mm sources in \citet[][]{Yamaguchi2019}, mostly owing to their smaller implied total infrared luminosities. This study does not report formal uncertainties or derive redshifts, which prohibits a more quantitative comparison. Bright SMGs beyond $z>4$ contribute $\sim1\times10^{-3}$ M$_{\odot}$ yr$^{-1}$ Mpc$^{-3}$ \citep{Swinbank2014, Michalowski2017}, about an order or magnitude lower than our best estimate. Results from fainter SMGs found in the deepest SCUBA surveys, with luminosities similar to \cone\ \citep[e.g.][]{Koprowski2017,Cowie2018} show a declining contribution at $2 < z < 5$, consistent with our estimates.

Interestingly, if \cone\ is representative, a population with similar properties could contribute as much to the CSFRD as all known ultraviolet-selected galaxies at similar redshifts combined.  This could even imply that dust-obscured star-formation continues to dominate the cosmic star-formation history beyond $z>4$ where current infrared-to-mm measurements are incomplete.

\subsection{Contribution to Stellar mass density}
{The high stellar mass $10^{10.8}$ M$_\odot$ and large space density $\sim3\times10^{-5}$ Mpc$^{-3}$ of  \cone\ imply a considerable cosmic stellar mass density (CSMD) in similar objects at $z\sim5$: $\rho^*=1.9^{+4.4}_{-1.5} \times 10^6$ M$_\odot$Mpc$^{-3}$, higher than reported for bright (S$_{850}> 4$ mJy) sub-mm galaxies \citep[$\approx0.5\times10^6$ M$_\odot$Mpc$^{-3}$;][]{Michalowski2017}. Comparing to the estimate of the CSMD based on HST-selected galaxies $\rho^*(z=5)=6.3\times 10^6$M$_\odot$Mpc$^{-3}$ \citep{Song2016}, suggests that they could contribute a significant fraction ($22^{+25}_{-16}$\%) to the total and perhaps even dominate the high-mass end. The high-mass end of the galaxy stellar mass function at high redshift $z>4$ is still uncertain and susceptible to biases. Current estimates for the number density of $\sim10^{10.8}$ M$_\odot$ galaxies at $z\sim5$ are $1-5\times10^{-5}$ Mpc$^{-3}$dex$^{-1}$ \citep{Duncan2014, Grazian2015, Song2016, Davidzon2017, Stefanon2017_900nm}, comparable to the space density derived for our ALMA-only galaxy \cone. Given that optical-IRAC dark galaxies are missing from these previous studies, it is therefore possible that about half the stellar mass density in high-mass galaxies at $z\sim5$ remains unaccounted for. }

\subsection{Implications for the formation of massive galaxies}\label{sec:massive}

Bright sub-millimeter-selected galaxies at $z>3$ with SFR$\gtrsim$1000 M$_{\odot}$yr$^{-1}$ are often hypothesized as progenitors of massive $z\sim2$ quiescent galaxies  \citep[e.g.][]{Toft2014,Spilker2018sci}. More recently, massive quiescent galaxies have been spectroscopically confirmed at $3<z<4$ \citep{Glazebrook2017,Schreiber2018qg}, but finding their progenitors at even earlier times is challenging. Generally, UV-selected galaxies at $z>4$ are presumed not abundant, massive, and star forming enough to produce the population of the earliest known massive quiescent galaxies with N$\sim3-5\times10^{-5}$ Mpc$^{-3}$ and Log (M/M$_\odot)\gtrsim10.6$ \citep[][]{Straatman2014}. 
Bright ($> 4$ mJy) sub-millimeter galaxies are a possible avenue, but it is difficult to establish a conclusive connection. Estimated SMG number densities at $z>4$ are low and uncertain $\sim0.1-3\times10^{-6}$ Mpc$^{-3}$ \citep{Ivison2016,Michalowski2017,Jin2018}. While their high SFRs indicate they will rapidly form massive galaxies, the modest inferred gas masses indicate gas depletion timescales on the order of $10-100$ Myr \citep[e.g.][]{Aravena2016a,Spilker2018sci}, and large duty cycle corrections are needed to make up for the low number densities. 

The inferred space densities and stellar mass of \cone\ on the other hand are already as large as those reported for the population of early quiescent galaxies at $3<z<4$ \citep[][]{Straatman2014,Nayyeri2014}. The large $\sim10^{11}M_\odot$ gas mass and modest SFR indicates much longer depletion timescales ($\sim200-500$ Myr), half the age of the Universe at this epoch, and implies a $\sim50-100$\% duty cycle for our adopted $4\lesssim z \lesssim6$ selection window.  \cone\ could therefore represent a more gradual path for massive galaxy growth compared to a rapid and bursty formation as has been found in some 
 bright merger-induced SMGs \citep[e.g.][]{Pavesi2018, Marrone2018}.
Overall, our results provide evidence for the existence of a sustained growth mode for massive galaxies in the early Universe.

Finally, the large systematic difference in SFR and stellar mass in particular depending on the assumed attenuation model for \ctwo\ suggests exercising caution when relating progenitors and descendants galaxies. Often such links are determined based on the capability of a progenitor population  to produce adequate numbers of sufficiently massive galaxies at later times \citep[e.g.,][]{Straatman2014,Toft2014,Williams2014,Williams2015}, or by the assumption that the rank-order on stellar mass can be reliably determined \citep[e.g.][]{Brammer2011,Behroozi2013cnd,Leja2013}. Such determinations could be more uncertain than has been accounted for if the derived stellar masses of individual galaxies are off by factors of $\sim8$ (see \S3.2). This is a particular concern at high redshift $z>3$, where very few galaxies have high enough SNR (sub-)mm observations for meaningful constraints on the dust attenuation model. These results are obviously based on only a single galaxy and larger samples with full optical-to-mm photometry are needed to determine the scatter in stellar mass.

\subsection{Future ALMA and JWST observations}

The dark nature of the \cone, non-detection in very deep stacked optical and near-IR data and the faint IRAC fluxes, suggests a prominent population of DSFGs at $z>4$. The high inferred redshift may point to the efficacy of blind surveys at $2-3$ mm to find the earliest dusty star forming galaxies \citep{Bethermin2015b,Casey2018b}. These galaxies are below the classical detection limit ($>4$ mJy) of bright galaxies found in single-dish sub-mm surveys. Deep 2 mm single-dish and wider $1-3$ mm ALMA (sub)-mm surveys are only just starting to push into this 
territory at very high redshift \citep[e.g.][]{Magnelli2019,Yamaguchi2019}. Until the launch of the {\it James Webb Space Telescope} ({\it JWST}), ALMA alone can find and study these galaxies.

Wide-area unbiased ALMA surveys covering 100's of arcmin$^2$ are necessary to further constrain their prominence in the early universe. Such surveys are feasible at $2-3$ mm with ALMA because of the relatively large size of the primary ALMA beam and the exquisite sensitivity to high redshift star forming galaxies even in short integration times \citep{Casey2019}. To date, none of the ALMA-only galaxies found has been spectroscopically confirmed, but spectral line scans for CO and in particular [CII] are efficient and feasible with ALMA. Future surveys with {\it JWST} will enable systematic studies of large samples of faint SMG galaxies. Large legacy surveys such as the
{\it JWST} Advanced Deep Extragalactic Survey (JADES) will likely characterize $\sim15-30$ galaxies similar to \cone\ (based on expected number densities from this work and \citealt{Zavala2018_3mm}, and observations described in \citealt{Williams2018}). {\it JWST} will have the capability to measure stellar population properties and redshifts, and in combination with ALMA far-infrared constraints, will enable a  detailed investigation into the star-formation, dust, and stellar population properties of massive galaxies in the early Universe.  

\vspace{0.5cm}

\acknowledgments

We are grateful to James Simpson for making part of S2COSMOS available for our analysis in advance of publication. We thank Jorge Zavala, Peter Behroozi and Pieter van Dokkum for helpful discussions. C.C.W. acknowledges support from the National Science Foundation Astronomy and Astrophysics Fellowship grant AST-1701546.  J.L. is supported by an NSF Astronomy and Astrophysics Postdoctoral Fellowship under award AST-1701487. This paper makes use of the following ALMA data: ADS/JAO.ALMA \#2018.1.01739.S, ADS/JAO.ALMA \#2015.1.00853.S,  ADS/JAO.ALMA \#2015.1.00861.S. ALMA is a partnership of ESO (representing its member states), NSF (USA), and NINS (Japan), together with NRC (Canada), NSC and ASIAA (Taiwan), and KASI (Republic of Korea), in cooperation with the Republic of Chile. The Joint ALMA Observatory is operated by ESO, AUI/NRAO, and NAOJ. The National Radio Astronomy Observatory is a facility of the National Science Foundation operated under cooperative agreement by Associated Universities, Inc.

\appendix

\begin{figure*}[th]
    \centering
    \includegraphics[width=0.8\textwidth]{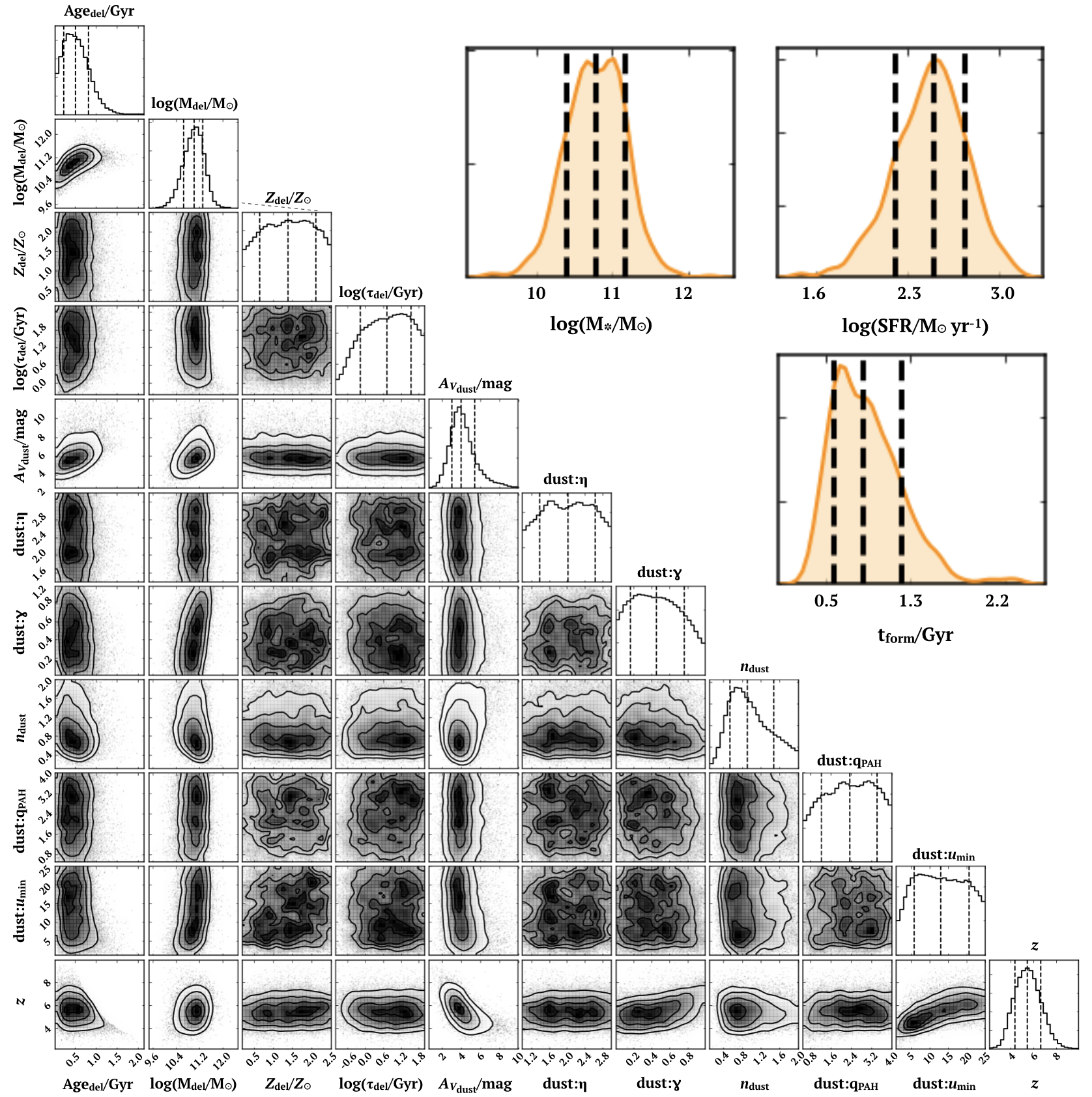}
    \caption{ Posterior probability distributions for the fitted parameters of \cone\ corresponding to the orange region in Figure \ref{sed}, assuming \citet{Charlot2000} dust attenuation with the slope left free. { M$_{\footnotesize\textrm{del}}$ refers to the total stars formed (integral of the delayed star formation history) whereas M$^{*}$ is the stellar mass excluding remnants and including mass loss due to stellar evolution, and is the parameter referred to throughout the text.} Parameters, their definitions, and their uncertainties are presented in Table \ref{tab:sedfit}. }
    \label{fig:pdf1}
\end{figure*}

\begin{figure*}[th]
    \centering
   \includegraphics[width=0.8\textwidth]{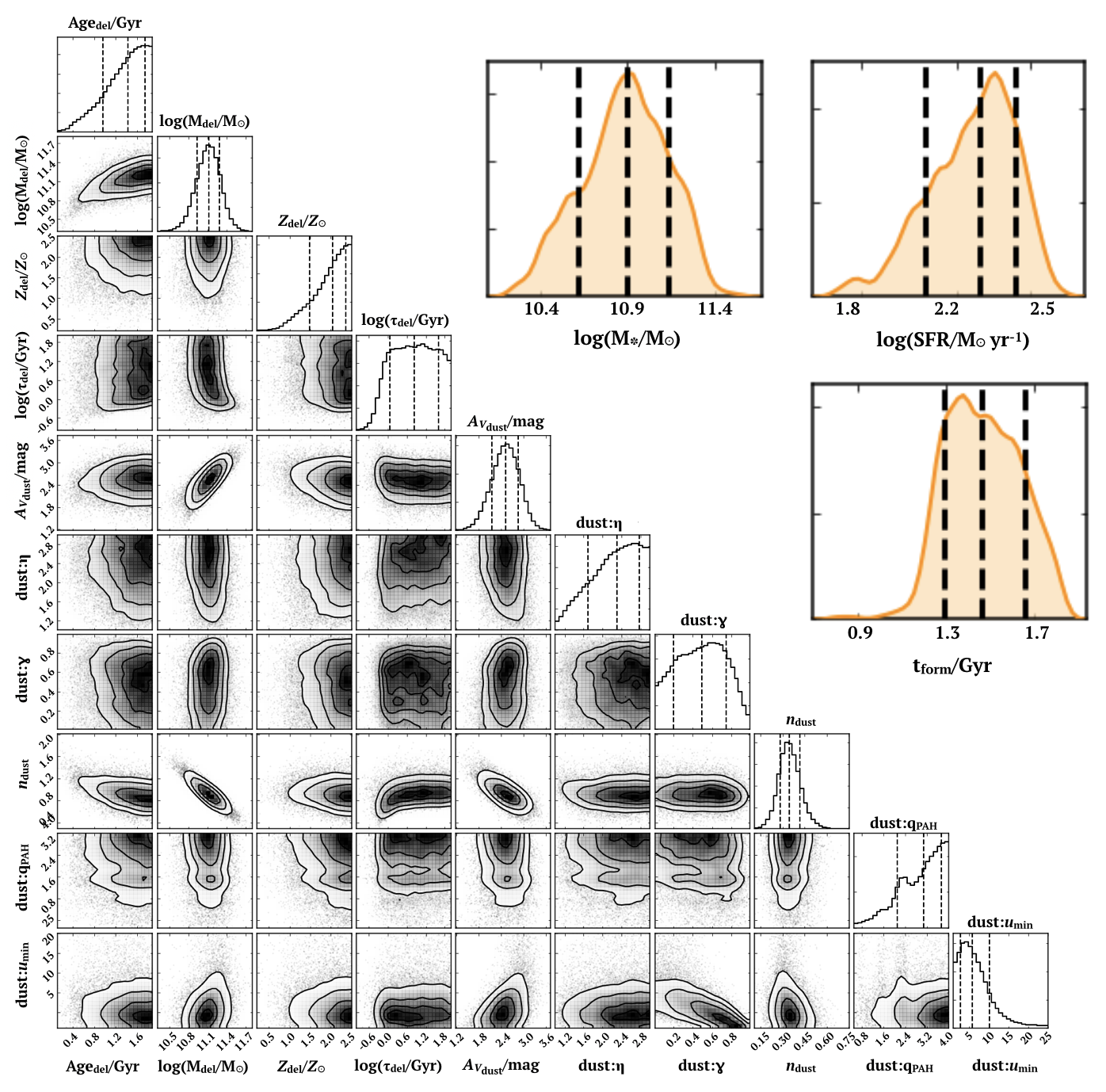}
     \caption{ Posterior probability distributions for the fitted parameters of \ctwo\ corresponding to the orange region in Figure \ref{sed}, assuming \citet{Charlot2000} dust attenuation with the slope left free, and fixed to the spectroscopic redshift. Measured parameters, their definitions and their uncertainties are presented in Table \ref{tab:sedfit}.  }
    \label{fig:pdf2}
\end{figure*}

{Here we present supplemental observational details for \cone\ and \ctwo. 
In Table \ref{tab:phot}, we present the deblended photometry (measurements and 1$\sigma$ uncertainties) based on the ALMA priors for the two 3-mm sources  that are used in the SED fitting analysis presented in Section \ref{sec:sed}. In Figures \ref{fig:pdf1} and \ref{fig:pdf2} we present the posterior probability distributions for the galaxy properties of both sources measured using the Bayesian SED fitting with \textsc{bagpipes}. }

\begin{deluxetable*}{lcccc}[t!]
\tablecaption{ALMA prior-based deblended photometry } 
\tablecolumns{5}
\tablewidth{0pt}
\tablehead{
\colhead{Band} &
\colhead{\cone\ flux density} &
\colhead{RMS } &
\colhead{\ctwo\ flux density } &
\colhead{RMS} \\
\colhead{} &
\colhead{[$\mu$Jy]} &
\colhead{ } &
\colhead{[$\mu$Jy]} &
\colhead{} \\
}
\startdata
SUBARU B & [ -1.3E-02 ] &  7.8E-03  &  --  &  -- \\ 
HSC g & [ 3.6E-03 ] &  1.5E-02  & [ 2.3E-02 ] &  1.1E-02 \\ 
SUBARU V & [ -7.6E-03 ] &  1.8E-02  & --  &  -- \\ 
HSC r & [ -1.2E-02 ] &  1.4E-02  &  9.3E-02  &  9.7E-03 \\ 
SUBARU rp & [ 1.2E-03 ] &  1.7E-02  &  --  &  -- \\ 
SUBARU ip & [ 2.0E-03 ] &  2.6E-02  &  --  &  -- \\ 
HSC i & [ -1.0E-02 ] &  2.1E-02  &  1.2E-01  &  1.5E-02 \\ 
HSC z & [ 1.2E-02 ] &  3.0E-02  &  1.3E-01  &  2.4E-02 \\ 
SUBARU zp & [ -1.5E-01 ] &  6.5E-02  &  --  &  -- \\ 
HSC Y & [ 2.2E-02 ] &  7.4E-02  & [ 4.9E-02 ] &  5.8E-02 \\ 
UltraVISTA Y & [ 9.7E-03 ] &  8.9E-02  & [ 1.6E-01 ] &  8.2E-02 \\ 
UltraVISTA J & [ -1.1E-01 ] &  1.0E-01  & [ 2.0E-01 ] &  8.9E-02 \\ 
UltraVISTA H & [ -5.9E-02 ] &  1.3E-01  &  6.3E-01  &  1.1E-01 \\ 
UltraVISTA Ks & [ 1.2E-01 ] &  8.8E-02  &  8.4E-01  &  8.7E-02 \\ 
\it{Spitzer}/IRAC 3.6$\mu$m & [ 3.2E-01 ] &  1.2E-01  &  1.7E+00  &  9.2E-02 \\ 
\it{Spitzer}/IRAC 4.5$\mu$m & [ 3.2E-01 ] &  1.3E-01  &  3.0E+00  &  9.9E-02 \\ 
\it{Spitzer}/IRAC 5.8$\mu$m & [ 5.2E+00 ] &  3.3E+00  &  8.8E+00  &  2.2E+00 \\ 
\it{Spitzer}/IRAC 8$\mu$m & [ 6.0E+00 ] &  4.1E+00  & [ 7.7E+00 ] &  2.9E+00 \\ 
\it{Spiter}/MIPS 24$\mu$m & [ 1.1E+01 ] &  1.5E+01  & [ 2.6E+01 ] &  1.5E+01 \\ 
\it{Herschel}/PACS 100$\mu$m & [ 7.1E+02 ] &  1.4E+03  & [ 1.5E+02 ] &  1.4E+03 \\ 
\it{Herschel}/PACS 160$\mu$m & [ -2.4E+03 ] &  3.2E+03  & [ -1.9E+03 ] &  3.2E+03 \\ 
\it{Herschel}/SPIRE 250$\mu$m & [ 3.7E+03 ] &  3.9E+03  & [ 1.0E+03 ] &  3.9E+03 \\ 
\it{Herschel}/SPIRE 350$\mu$m & [ 4.3E+03 ] &  4.9E+03  & [ -5.0E+02 ] &  4.9E+03 \\ 
\it{Herschel}/SPIRE 500$\mu$m & [ 5.5E+03 ] &  5.8E+03  & [ 4.4E+03 ] &  5.8E+03 \\ 
SCUBA2 850$\mu$m &  3.5E+03  &  1.1E+03  & [ 4.0E+02 ] &  1.1E+03 \\ 
ALMA 3mm &  1.6E+02  &  2.2E+01  &  7.5E+01  &  1.0E+01 \\ 
VLA 3 GHz &  1.0E+01  &  2.4E+00  & [ 3.9E+00 ] &  2.4E+00 \\ 
VLA 1.4 GHz & [ 1.7E+01 ] &  1.7E+01  & [ 4.0E+01 ] &  1.7E+01 \\ 
\enddata
\tablecomments{Subaru optical photometry for \ctwo\ are not measured owing to cosmetic defects in the mosaics that cross the location of the source. Photometric upper limits as indicated by downward arrows in Figure \ref{sed} are at the RMS values of photometric points in this table where S/N$<$1.  Non-significant photometric points (with SNR$<3$) are indicated with brackets. }\label{tab:phot}
\end{deluxetable*}

\bibliographystyle{aasjournal}
\bibliography{ApJL_template}

\end{document}